\documentclass[conference]{IEEEtran}
  \usepackage[pdftex]{graphicx}
\usepackage{amsmath,amssymb,amsthm, latexsym}
\usepackage[font=small,labelfont=bf]{caption}
\usepackage{hyperref}

\usepackage[english]{babel}
\usepackage[utf8]{inputenc}
\usepackage{algorithm}
\usepackage[noend]{algpseudocode}
\usepackage{xcolor}
\usepackage[super]{nth}
\renewcommand\thesection{\Roman{section}} 
\renewcommand\thesubsectiondis{\thesection.\Alph{subsection}}
\renewcommand\thesubsubsectiondis{\thesubsectiondis.\arabic{subsubsection}}

\begin{document}
%
\title{Bit-permuted coded modulation for polar codes}
\author{\IEEEauthorblockN{Saurabha R. Tavildar}
\IEEEauthorblockA{Email: tavildar at gmail }
}

\thispagestyle{plain}
\pagestyle{plain}
\maketitle
\begin{abstract}
~We consider the problem of using polar codes with higher order modulation over AWGN channels. Unlike prior work, we focus on using modulation independent polar codes. That is, the polar codes are not re-designed based on the modulation used. Instead, we propose bit-permuted coded modulation (BPCM): a technique for using the multilevel coding (MLC) approach for an arbitrary polar code. The BPCM technique exploits a natural connection between MLC and polar codes. It involves applying bit permutations prior to mapping the polar code to a higher order modulation. The bit permutations are designed, via density evolution, to match the rates provided by various bit levels of the higher order modulation to that of the polar code. 

~We demonstrate performance of the BPCM technique using link simulations and density evolution for the AWGN channel. We compare the BPCM technique with the bit-interleaved coded modulation (BICM) technique. When using polar codes designed for BPSK modulation, we show gains for BPCM over BICM with random interleaver of up to $0.2$ dB, $0.7$ dB and $1.4$ dB for $4$-ASK, $8$-ASK, and $16$-ASK respectively.
\end{abstract}

\section{Introduction}
Polar codes, introduced in \cite{Arikan}, were proved to achieve the symmetric capacity for BDMCs. Following \cite{Arikan}, works in \cite{MT}, \cite{Trifonov} have considered use of polar codes for AWGN channel focusing on the BPSK modulation. In this paper, we focus on use of polar codes over AWGN channels with higher order modulation. This has been considered in various works in \cite{Seidl}, \cite{HM}, \cite{SLY}. Our work builds on this prior work. The key contribution of our work is the use of modulation independent polar codes. Our work is discussed in Sections~\ref{sec:mipc} and \ref{sec:misc}. We first  discuss the prior work in \hyperref[sec:prior_work]{Section~\ref*{sec:prior_work}}. 	

\section{Overview of prior work}\label{sec:prior_work}

\subsection{Coded Modulation}

Coded modulation deals with mapping a binary code to a higher order modulation. The higher order modulations typically considered for AWGN channel are amplitude shift keying (ASK) or pulse amplitude modulation. Typically, 4-ASK, 8-ASK, 16-ASK are considered. For complex channels, such as complex baseband representations of wireless channels, modulations such as 8-PSK, 16-QAM, 64-QAM, and 256-QAM are used. These modulations are defined by input alphabet $X$ that is typically normalized to have a unit power. In this paper, we focus on the AWGN channel and hence consider ASK modulations. 

Examples of coded modulation are trellis-coded modulation \cite{Ung87}, the multilevel coding (MLC) \cite{IH77} and bit-interleaved coded modulation (BICM) \cite{Zehavi}, \cite{Caire}. The BICM technique has been design choice of standards such as LTE \cite{212} due to its simplicity and close to optimal performance.

\subsection{Polar coding for higher order modulation}

The work in \cite{Seidl} \cite{HM} \cite{SLY} discussed application of these coded modulation techniques in the context of polar codes. A summary of work on polar coded modulation is given in \cite{ArikanMLC}. The work in \cite{Seidl} presented a unified approach for designing polar codes for higher order modulation for the AWGN channel. This is summarized next. 

Let $W: X \rightarrow Y$ be channel with input alphabet of size $2^k$. The key idea in \cite{Seidl} was to transform channel $W$ into $k$ binary bit channels, and design polar codes for the bit channels. Two transformation techniques were presented: serial binary partition (SBP) and parallel binary partition (PBP). 

The SBP transform ($\phi$) is closely related to the MLC technique. It transforms channel $W$ into $k$ bit channels
$\{ B_{\phi}^{(0)}, \cdots,  B_{\phi}^{(k-1)} \}$. It is defined by labeling rule $L_{\phi}$ that bijectively maps $k$ bits to the input $X$.
\begin{eqnarray}\label{eq:label}
L_{\phi}: [b_0, \cdots, b_{k-1}] \rightarrow x \in X
\end{eqnarray}  
The $B_{\phi}^{(i)}$ are bit channels with input $b_i$ that have access to output $Y$ as well as to $b_j$ for $j < i$: 
\begin{eqnarray}\label{eq:sbp1}
B_{\phi}^{(i)}: \{ 0, 1 \} & \rightarrow & Y \times \{ 0, 1 \}^i,
\end{eqnarray}
and have capacity: 
\begin{eqnarray}\label{eq:sbp2}
I ( B_{\phi}^{(i)}; Y) & = & I(B_i; Y | B_0, \cdots, B_i).
\end{eqnarray}
Equations~(\ref{eq:sbp1}) and (\ref{eq:sbp2}) assume an MLC approach since bit channel $B_{\phi}^{(i)}$ has access to inputs of channels $B_{\phi}^{(j)}$ for $j < i$.

The PBP transform ($\bar{\phi}$) is closely related with BICM technique and maps $W$ to $k$ \textit{independent} binary input channels. Similar to \hyperref[eq:label]{Equation~\ref*{eq:label}}, the PBP is defined by a labeling rule that bijectively maps $k$ bits to the input $X$. Thus, $B_{\bar{\phi}}^{(i)}: \{ 0, 1 \} \rightarrow Y$ are bit channels with capacity  $I(B_i; Y)$. The bit channel $B_{\bar{\phi}}^{(i)}$ does not have access to inputs of other channels $B_{\bar{\phi}}^{(j)}$.

\begin{figure}[!ht]
\vspace*{-0.15in}
\includegraphics[width=\linewidth]{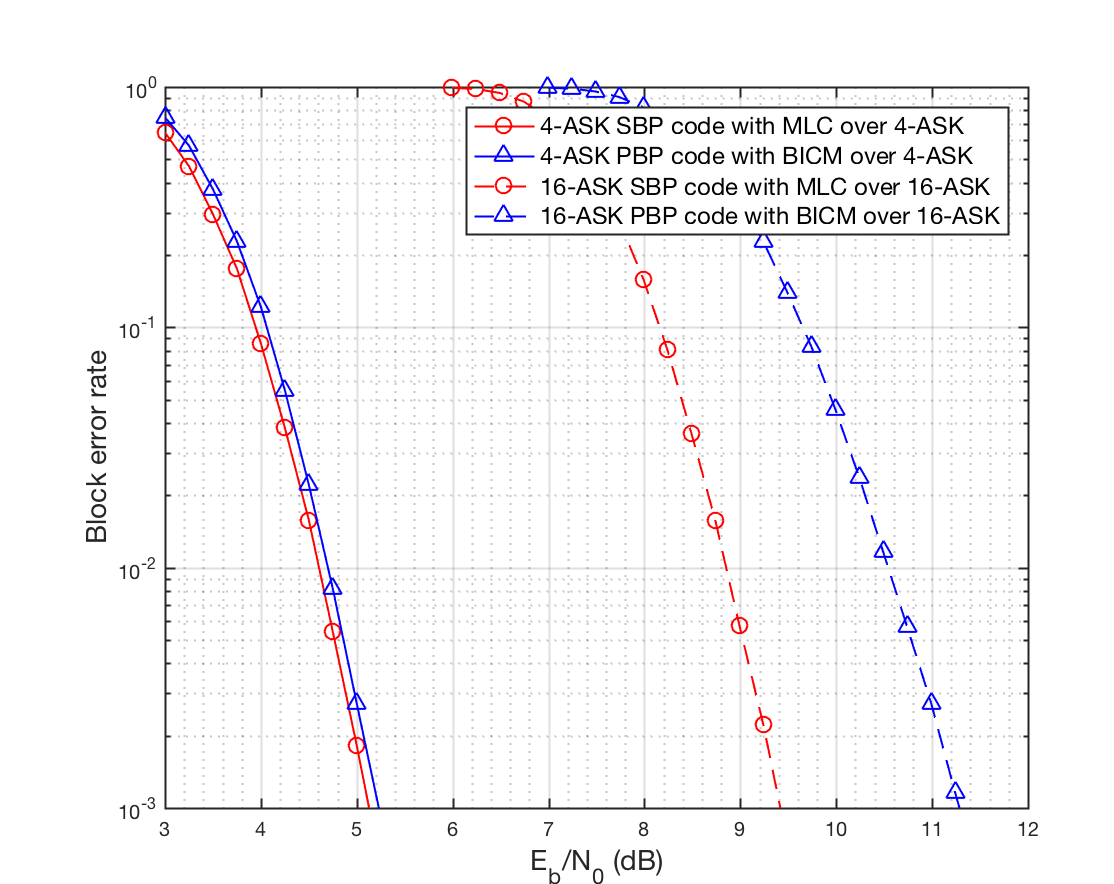}
\caption{4-ASK and 16-ASK AWGN link sim; N=1024, rate=$\frac{1}{2}$.}\label{fig:4_16_ask_prior}
\end{figure}

\begin{figure}[!ht]
\vspace*{-0.15in}
\includegraphics[width=\linewidth]{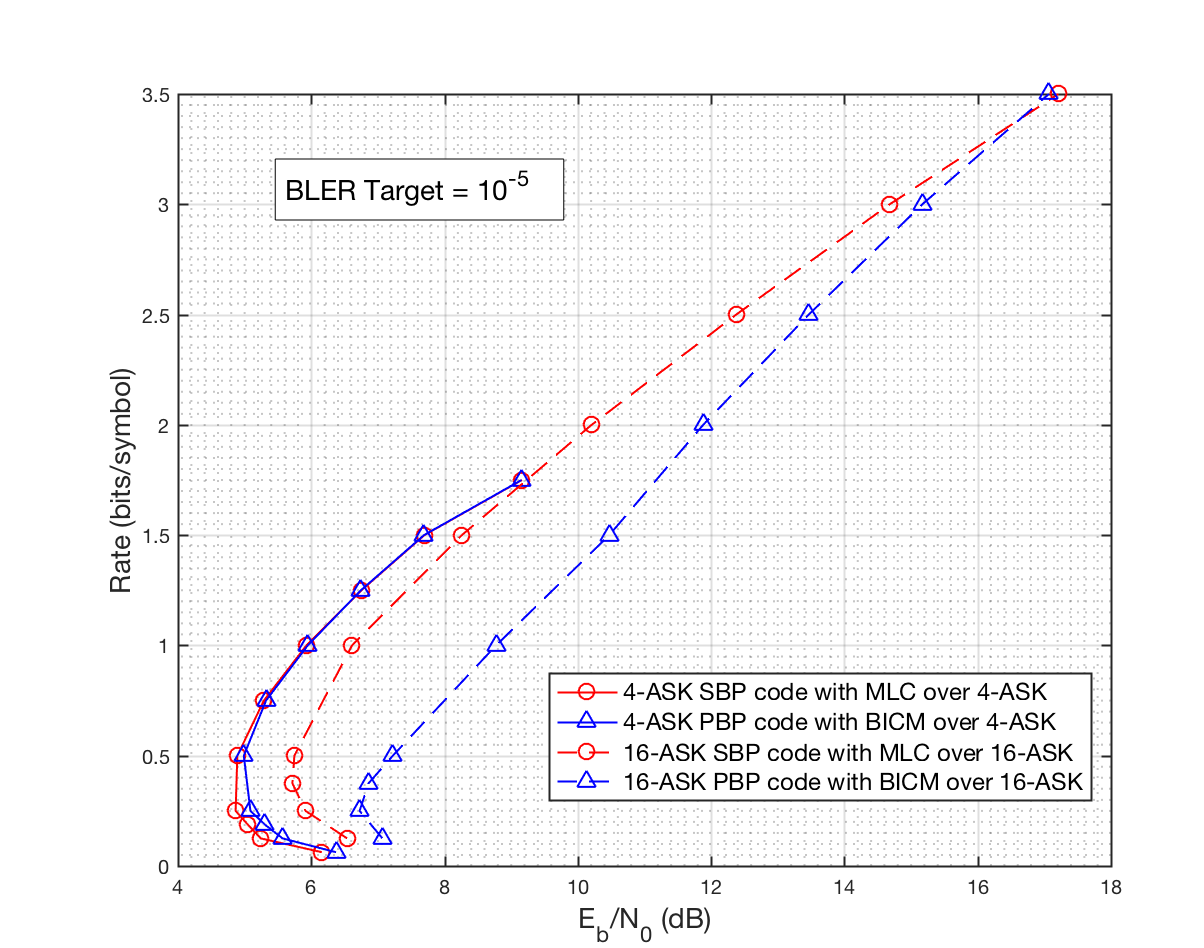}
\caption{4-ASK and 16-ASK AWGN GA-DE; N = 1024.}\label{fig:4_16_ask_de}
\vspace*{0.05in}
\end{figure}

Both SBP and PBP are analyzed in detail in \cite{Seidl}. It is shown that labeling rule significantly impacts the performance. Two labeling rules considered are set partitioning (SP) labeling (see \cite{Ung87}) and Gray labeling (see \cite{Gray}). It is shown in \cite{Seidl} that:
\begin{itemize}
\setlength\itemsep{0.25em}
\item for SBP, SP labeling outperforms Gray labeling;
\item for PBP, Gray labeling outperforms SP labeling;
\item SBP with SP labeling has the best performance.
\end{itemize}
Based on these conclusions, throughout this paper, SBP is used with SP labeling and PBP is used with Gray labeling. We present the results summarizing the prior work below. \hyperref[fig:4_16_ask_prior]{Figure~\ref*{fig:4_16_ask_prior}} shows AWGN link simulations with Monte-Carlo code constructions, and succssive cancellation decoder (SCD) for a rate $1/2$ code. \hyperref[fig:4_16_ask_de]{Figure~\ref*{fig:4_16_ask_de}}shows results using the Gaussian approximated density evolution (GA-DE) technique for code construction and BLER estimation for SCD for a range of rates. For both PBP and SBP, a binary input AWGN channel of the same capacity as the bit channels is assumed for GA-DE computation. Gaussian approximation introduces inaccuracy for higher order modulation especially for Gray labeling. To get more accurate performance for PBP, first order polarized bit channel capacities are calculated numerically as described in \cite{Seidl}.  We note that for the PBP/BICM approach, the original construction in \cite{Seidl} is used.

In \cite{HM}, an approach of compound polar codes is presented. This appraoch is closely related to the PBP approach of \cite{Seidl}. In our simulations, the performance of PBP in \cite{Seidl} was similar to that of compound polar codes in \cite{HM}. In \cite{SLY}, an approach of designing an interleaver is considered. This approach is discussed further in \hyperref[app:bicm]{Apppendix~\ref*{app:bicm}}.

\subsection{Limitations of prior work}

One limitation of prior work is the modulation dependence of polar codes. In prior work, the frozen indices of the polar code are selected based on the modulation being used. In some systems, this is not desirable. For example, in LTE \cite{212}, an interleaver is used to map the same code to different modulations. One reason for selecting a modulation independent code is the hybrid-ARQ (H-ARQ) protocol. For H-ARQ, if the receiver is to be able to soft combine multiple transmissions, then the underlying code needs to be the same. Thus modulation dependent polar codes would require all H-ARQ transmissions to use the same modulation.  Motivated by this, we consider modulation independent polar coding next. 

\section{Modulation independent polar coding}\label{sec:mipc}

We propose a technique for mapping an arbitrary polar code to a higher order modulation. Initially, for ease of exposition, we focus on 4-ASK in Sections~\ref{sec:polar_mlc} to \ref{sec:4ask}. The extension to 16-ASK is natural and discussed afterwards in \hyperref[sec:16ask]{Section~\ref*{sec:16ask}}. For 8-ASK, since the number of bit levels is not a power of 2, special consideration is needed. A solution for 8-ASK is discussed in \hyperref[sec:8ask]{Section~\ref*{sec:8ask}}. 

\subsection{MLC interpretation of arbitrary polar code over 4-ASK}\label{sec:polar_mlc}

Based on the results of \cite{Seidl} that the MLC approach outperfoms the BICM approach, we consider whether the MLC approach can be generalized without redesigning the code. We note that a connection between the MLC technique and polar codes has been observed in \cite{Seidl} and \cite{ArikanMLC}. The observation is that differing capacities of bit levels of a higher order modulation can be viewed as providing a natural polarization. Here, we make another observation connecting the MLC technique and polar codes. We observe that any polar code can be viewed as a multilevel code. 

\begin{figure}[!ht]
\includegraphics[width=\linewidth]{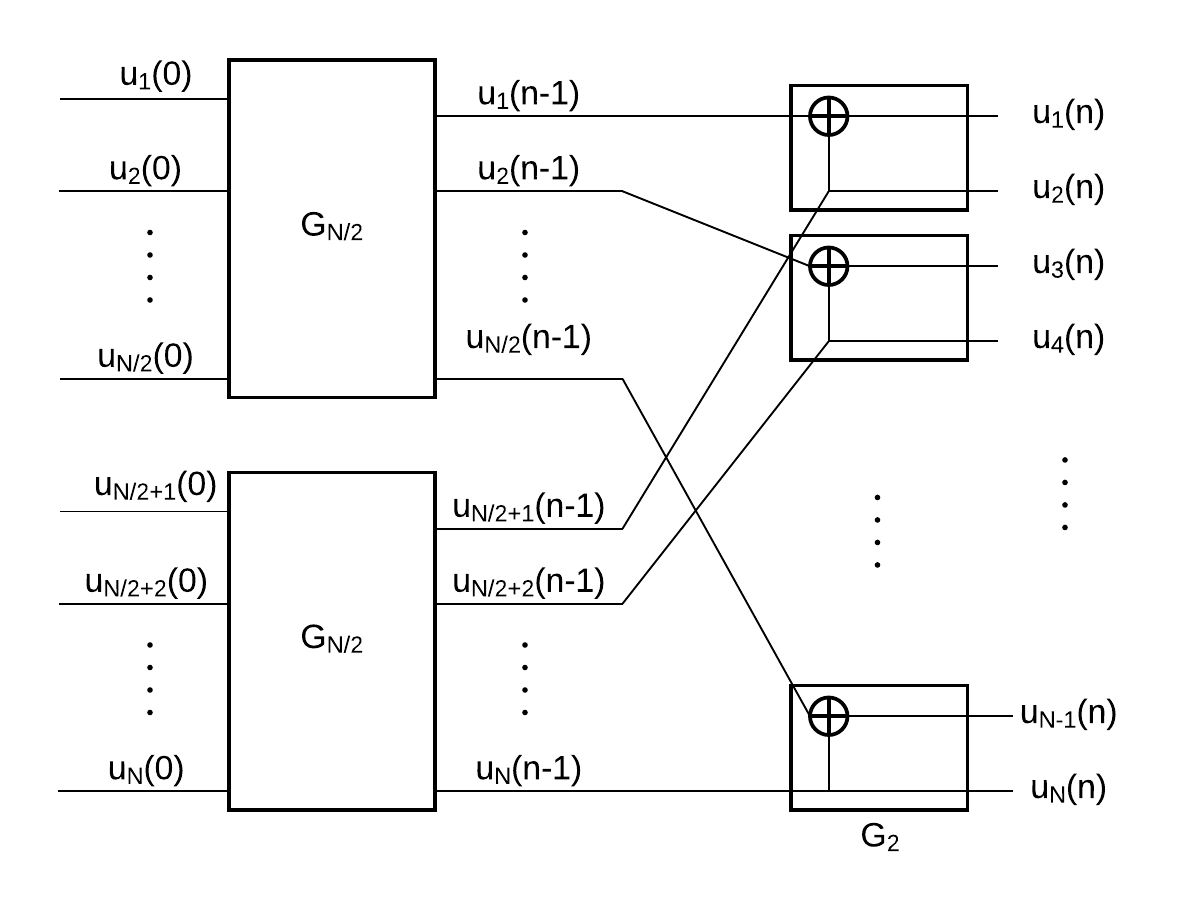}
\caption{A recursive representation of polar transform $G_N$}\label{fig:polar_mapping}
\end{figure}

Consider the polar transformation as shown in \hyperref[fig:polar_mapping]{Figure~\ref*{fig:polar_mapping}}. The figure shows the transformation:
\begin{eqnarray*}
\{u_i(0)\}_{i=1}^{i=N} & \rightarrow & \{u_i(n)\}_{i=1}^{i=N}, ~ n = \log_2{N}
\end{eqnarray*}
 where $\{u_i(0)\}_{i=1}^{i=N}$ denote the information bits (including frozen bits) and $\{u_i(n)\}_{i=1}^{i=N}$ denote the coded bits.   \hyperref[fig:polar_mapping]{Figure~\ref*{fig:polar_mapping}} shows the polar transformation can be represented recursively. In particular, we consider two subcodes:
\begin{itemize}
\setlength\itemsep{0.5em}
\item $\{u_i(n\!-\!1)\}_{1}^{\frac{N}{2}}$ is a function of $\{u_i(0)\}_{1}^{\frac{N}{2}}$. 
\item $\{u_i(n\!-\!1)\}_{\frac{N}{2}+1}^{N}$ is a function of $\{u_i(0)\}_{\frac{N}{2}+1}^{N}$. 
\end{itemize} 
So any polar code can be viewed as a multilevel code with two component codes and the mapping as below:
\begin{eqnarray*}
\text{\nth{1} subcode}:&&\{u_i(0)\}_{1}^{\frac{N}{2}} \rightarrow  ~\{u_i(n\!-\!1)\}_{1}^{\frac{N}{2}} \\ \vspace*{0.1in}
\text{\nth{2} subcode}:&&\{u_i(0)\}_{\frac{N}{2}+1}^{N} \rightarrow  ~\{u_i(n\!-\!1)\}_{\frac{N}{2}+1}^{N} \\ \vspace*{0.1in}
\text{4-ASK mapping}:&& (u_i(n\!-\!1), u_{i + \frac{N}{2}}(n\!-\!1)) \rightarrow  X_i, 
\end{eqnarray*} 
where $X_i$ is the $\text{i}^{\text{th}}$ modulation symbol. One scheme to map the subcodes to 4-ASK is:
\begin{itemize}
\setlength\itemsep{0.25em}
\item Map \nth{1} subcode to the \nth{1} bit level of 4-ASK;
\item Map \nth{2} subcode to the \nth{2} bit level of 4-ASK.
\end{itemize}


The main problem with this scheme is that the rates of the subcodes are dictated by polarization of the channel for which the polar code was designed. These rates do not necessarily match the modulation polarization which is the polarization of the bit levels of 4-ASK. This creates a mismatch between code polarization and modulation polarization. In the context of using polar codes designed for the BPSK channel over 4-ASK, the modulation polarization of 4-ASK with SP labeling appears to be higher than the polarization provided by polarizing two BPSK channels. To resolve this mismatch, we propose the bit-permuted coded modulation (BPCM) technique.

\subsection{Motivation for BPCM}

The key observation behind BPCM: for a given labeling rule, for 4-ASK, there are two capacity preserving ways to recover modulation capacity. These two ways lead to different bit channel capacities. This is done by changing the order of bits as demonstrated in equations below:
\begin{eqnarray*}
I(X; Y) & = & I(B_0; Y) + I(B_1; Y|B_0)~~(P_1), \\
 & = & I(B_1; Y) + I(B_0; Y|B_1)~~(P_2).
\end{eqnarray*}
Here, permutation $P_1$ refers to the conventional order of decoding the bits whereas permutation $P_2$ refers the reverse order. These two approaches have differing bit capacities as shown in \hyperref[table4:ask_cap]{Table~\ref*{table:4ask_cap}}.
\begin{table}[!h]
\begin{center}
\begin{tabular}{|c|c|c|c|c|}
\hline
  & $I(X; Y)$ &  $I(B_{P(0)}; Y)$  & $I(B_{P(1)}; Y|B_{P(0)})$ \\ \hline
$P = P_1$ & 1.34 & 0.39 & 0.95 \\ \hline
$P = P_2$ & 1.34 & 0.78 & 0.56 \\ \hline
\end{tabular}
\caption[4-ASK ]{4-ASK bit channel capacities for $P_1$ and $P_2$ at 5 dB}
\label{table:4ask_cap}
\end{center}
\end{table} 

It can be seen from \hyperref[table:4ask_cap]{Table~\ref*{table:4ask_cap}} that permutation $P_2$ is less polarizing than permutation $P_1$. This along with the empirical observation that $P_1$ is more polarizing than polarization of two BPSK channels suggests use of permutation $P_2$ to resolve the mismatched polarization. The BPCM technique involves using either permutation $P_1$ or permutation $P_2$ for each modulation symbol. For symbols that use permutation $P_2$, coded bits from the \nth{1} subcode are mapped to the \nth{2} bit level and coded bits from the \nth{2} subcode are mapped to the \nth{1} bit level of the 4-ASK modulation. This can help resolve the mismatch between code polarization and modulation polarization if the bit permutations are designed carefully. The design of bit permutations, that is determining which permutation is used for each of the modulation symbols is discussed in \hyperref[sec:design_bpcm]{Section~\ref*{sec:design_bpcm}}. Before that, we summarize the BPCM architecture in \hyperref[sec:arch_bpcm]{Section~\ref*{sec:arch_bpcm}}.

\subsection{The BPCM architecture}\label{sec:arch_bpcm}

\begin{figure}[!ht]
\vspace*{-0.15in}
\includegraphics[width=\linewidth]{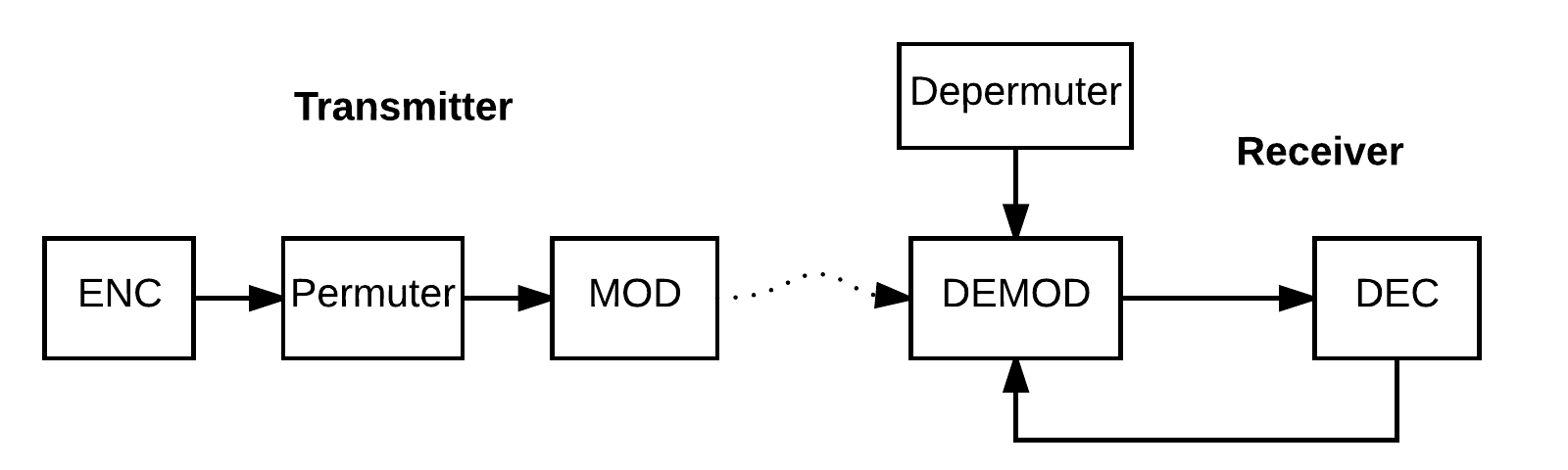}
\caption{BPCM architecture}\label{fig:bpcm}
\vspace*{0.05in}
\end{figure}

\begin{figure}[!ht]
\vspace*{-0.15in}
\includegraphics[width=\linewidth]{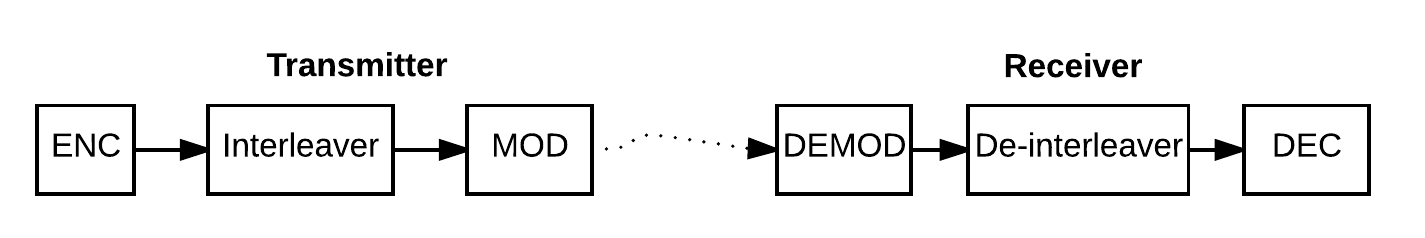}
\caption{BICM architecture}\label{fig:bicm}
\vspace*{0.05in}
\end{figure}

\begin{figure}[!ht]
\includegraphics[width=\linewidth]{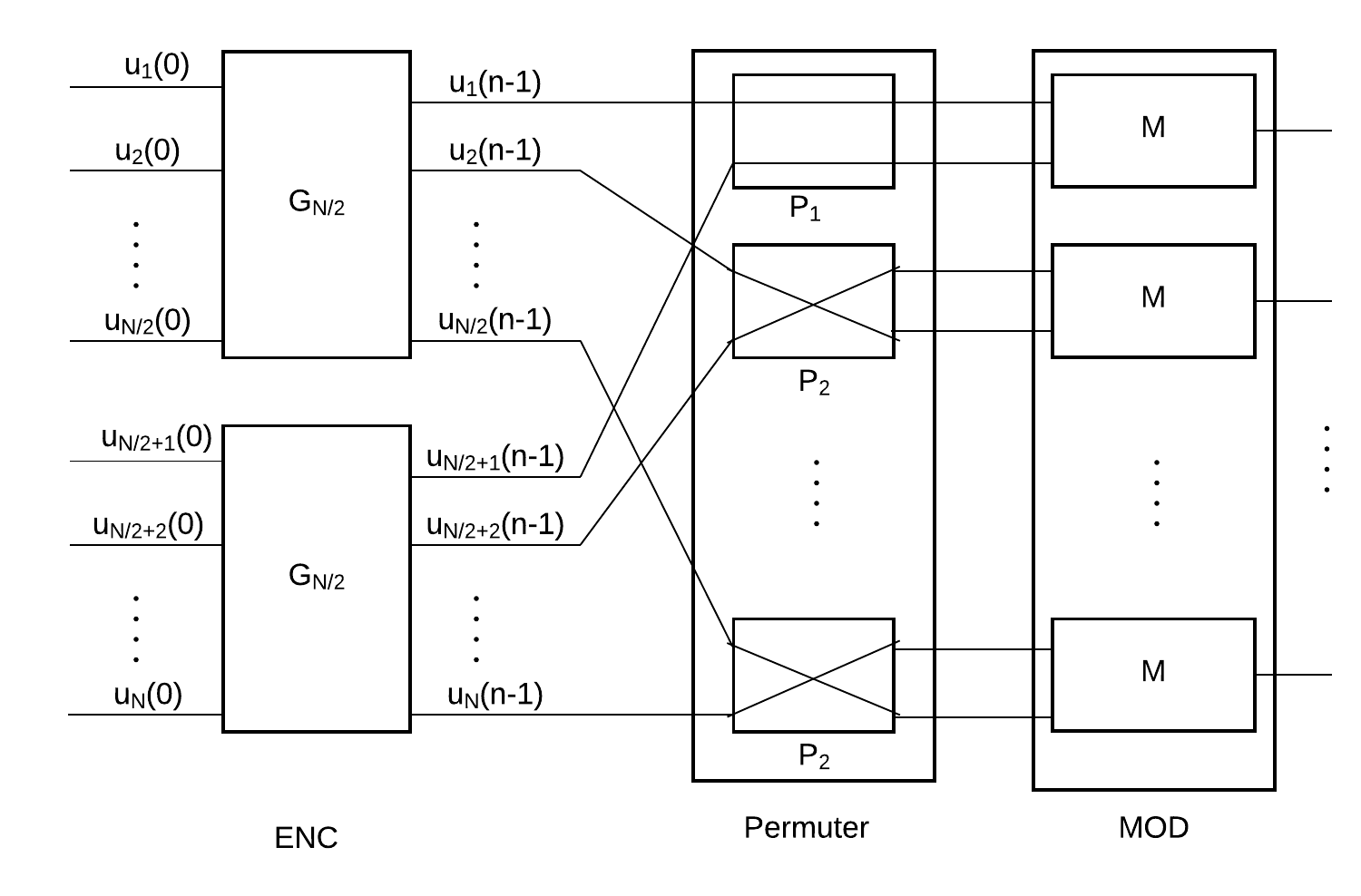}
\caption{A BPCM example for 4-ASK}\label{fig:polar_BPCM}
\vspace*{0.05in}
\end{figure}

\hyperref[fig:bpcm]{Figure~\ref*{fig:bpcm}} shows the BPCM architecture which can be contrasted with the traditional BICM architecture shown in \hyperref[fig:bicm]{Figure~\ref*{fig:bicm}} . On the transmitter side, BPCM uses a permuter in contrast to an interleaver in BICM. On the receiver side, BPCM is an MLC architecture requiring feedback from the decoder to the demodulator.  The decoder feeds back the decoded subcodes to the demodulator. The demodulator uses the feedback and permutations to compute the LLRs for the next subcode. \hyperref[fig:polar_BPCM]{Figure~\ref*{fig:polar_BPCM}} shows the transmitter architecture in BPCM in more detail. In particular it shows the permuter block comprising of a number of permutation operations, one for each modulation symbol. Each of the modulation symbol uses one of two permutations: $P_1$ or $P_2$. 

\subsection{Design of bit permutations}\label{sec:design_bpcm}

We use GA-DE to design the bit permutations. The design algorithm assumes permutation $P_1$ for all modulation symbols as the starting point. Then block error rate (BLER) estimated via GA-DE technique is used to determine the best permutation for each symbol sequentially. This is repeated for multiple iterations over the block length of the code. We typically see convergence in less than 10 iterations. This is described in \hyperref[alg:bpcm]{Algorithm~\ref*{alg:bpcm}} below. BLER evaluation is  based on GA-DE without re-selecting the set of frozen bits for the polar code.

\begin{algorithm}
\caption{BPCM permutation design}\label{alg:bpcm}
\begin{algorithmic}[1]
\Procedure{Permute}{$N, k, snr$}\Comment{N = block length}
\State $N_s \gets N/k$ \Comment{$N_s=$ \# of modulation symbols}
\State $P \gets \text{ones}(N_s)$
\State $M_k \gets k!$ \Comment{$M_k=2$ for 4-ASK}
\For{$ iter = 1 : 10 $} \Comment{Max 10 iterations}
\For{$ i = 1 : N_s $}
\State $E \gets \text{zeros}(M_k)$ 
\For{$ l = 1 : M_k$}
\State $P(i) = l$
\State $E[l] = \textsc{Estimate Bler}(P, snr)$
\EndFor
\State $P(i) = \text{argmin}(E)$
\EndFor
\EndFor
\State \textbf{return} $P$
\EndProcedure
\end{algorithmic}
\end{algorithm}

\subsection{Simulation results for 4-ASK}\label{sec:4ask}

\begin{figure}[!h]
\vspace*{-0.2in}
\includegraphics[width=\linewidth]{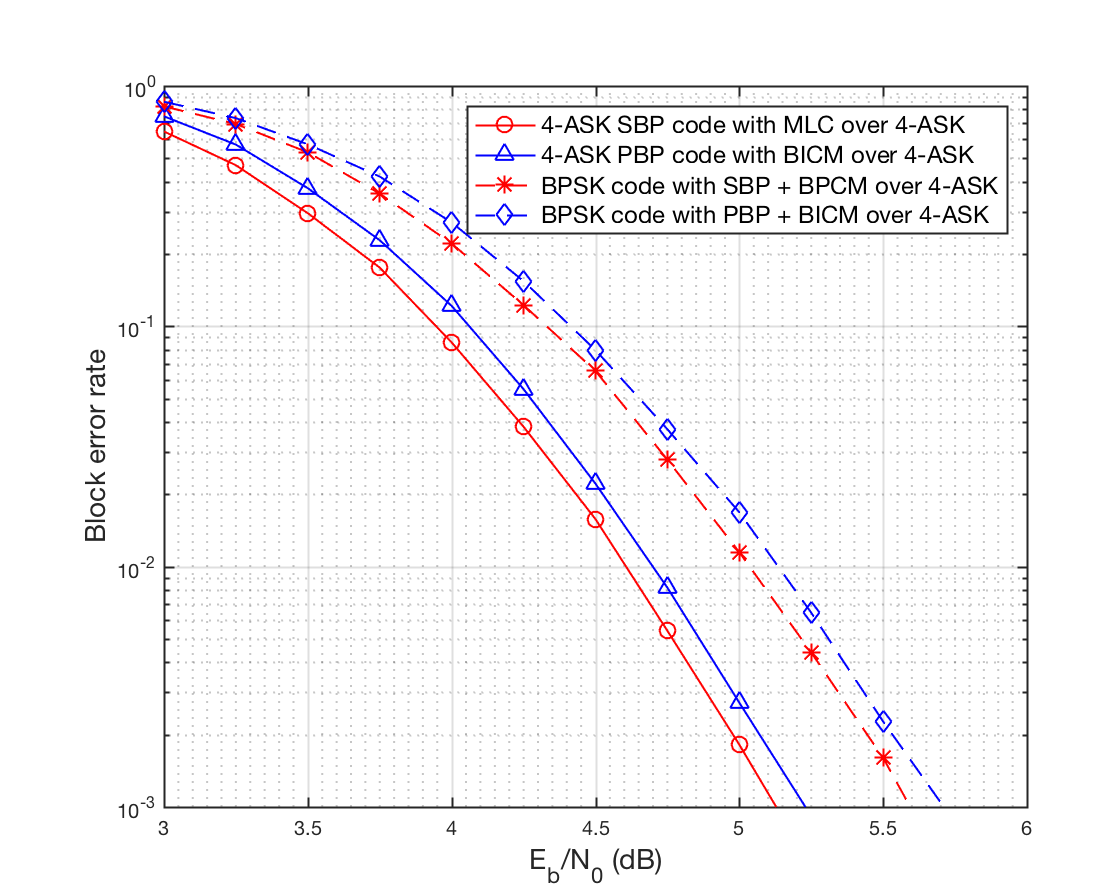}
\caption{4-ASK AWGN link sim; N=1024, rate=$\frac{1}{2}$.}\label{fig:4ask_bler_new}
\vspace*{0.05in}
\end{figure}

\begin{figure}[!ht]
\vspace*{-0.1in}
\includegraphics[width=\linewidth]{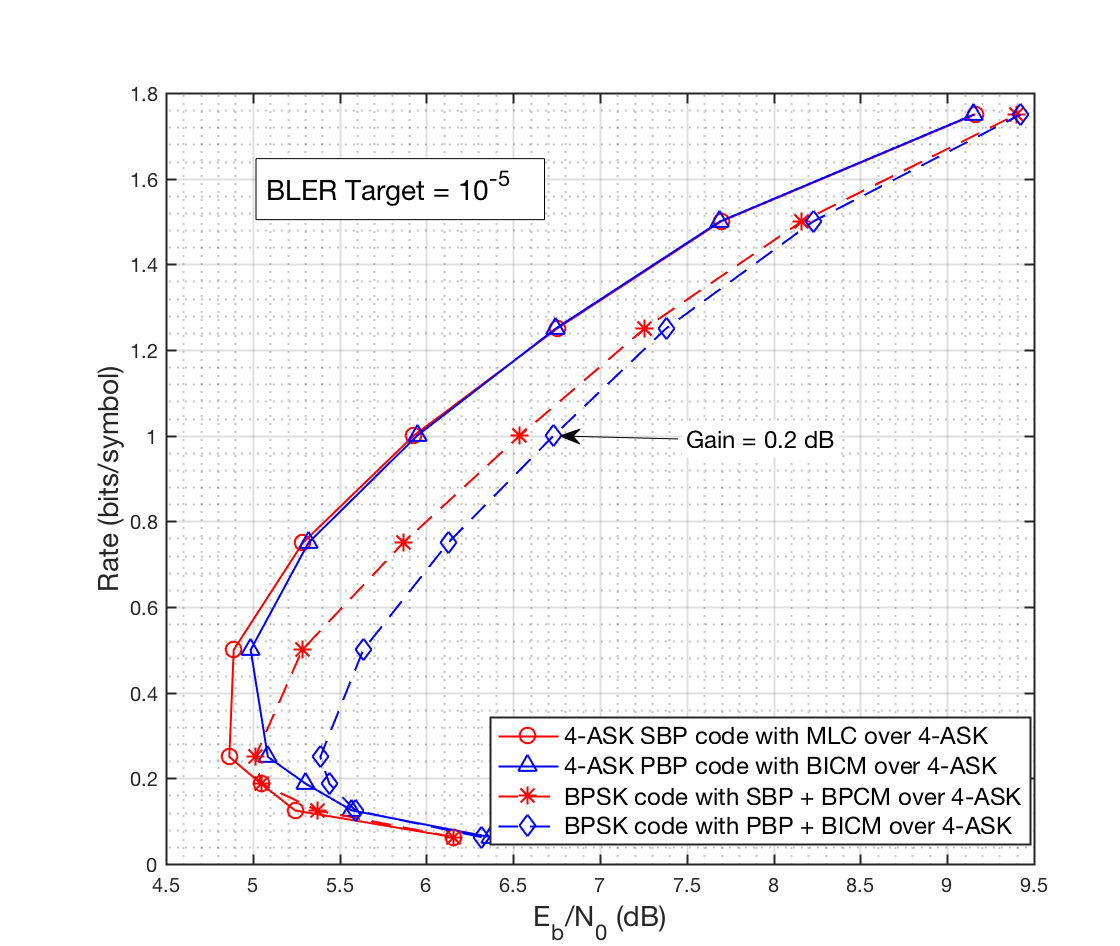}
\caption{4-ASK GA-DE; N = 1024.}\label{fig:4ask_de_new}
\vspace*{0.05in}
\end{figure}

We present simulation results for the BPCM technique. To illustrate the BPCM technique, polar codes designed for BPSK modulation (called BPSK codes) are used over 4-ASK. The BPCM technique is compared with with BICM technique for BPSK codes as well as modulation dependent codes. The BPSK code used for the link simulation was designed via Monte-Carlo simulation at $2.5$ dB $E_b/N_0$. For GA-DE results, the BPSK codes were designed for the lowest SNR where the code achieves the BLER target of $10^{-5}$. For using BPSK codes with BICM technique, a random interleaver is used, and bit-channel capacities of the modulation with Gray labeling are used for the computation. No interleaver is used for modulation dependent PBP codes, and no permutation is used for modulation dependent SBP codes. \hyperref[fig:4ask_bler_new]{Figure~\ref*{fig:4ask_bler_new}} shows AWGN link simulation results. \hyperref[fig:4ask_de_new]{Figure~\ref*{fig:4ask_de_new}} shows AWGN GA-DE results. 

The main results in the rate regime of interest are: 
\begin{itemize}
\setlength\itemsep{0.25em}
\item For BPSK codes, BPCM bests BICM by up to $0.2$ dB;
\item BPSK codes with BPCM is worse than modulation specific SBP or PBP codes by up to $0.6$ dB.
\end{itemize}

\subsection{Design and results for 16-ASK}\label{sec:16ask}

The differences between 4-ASK and 16-ASK BPCM are:
\begin{itemize}
\setlength\itemsep{0.25em}
\item 4 subcodes for 4 different bit levels of 16-ASK are used;
\item the 4 subcodes for an arbitrary polar code are:	
\begin{equation*}
\{u_i(0)\}_{1 + j \frac{N}{4} }^{\frac{N}{4} + j\frac{N}{4}}  \rightarrow \{u_i(n\!-\!2)\}_{1 + j \frac{N}{4} }^{\frac{N}{4} + j\frac{N}{4}}, 0 \leq j \leq 3 ;
\end{equation*}
\item $M_k = 24$ permutations used for \hyperref[alg:bpcm]{Algorithm~\ref*{alg:bpcm}}.
\end{itemize}

\begin{figure}[!ht]
\vspace*{-0.05in}
\includegraphics[width=\linewidth]{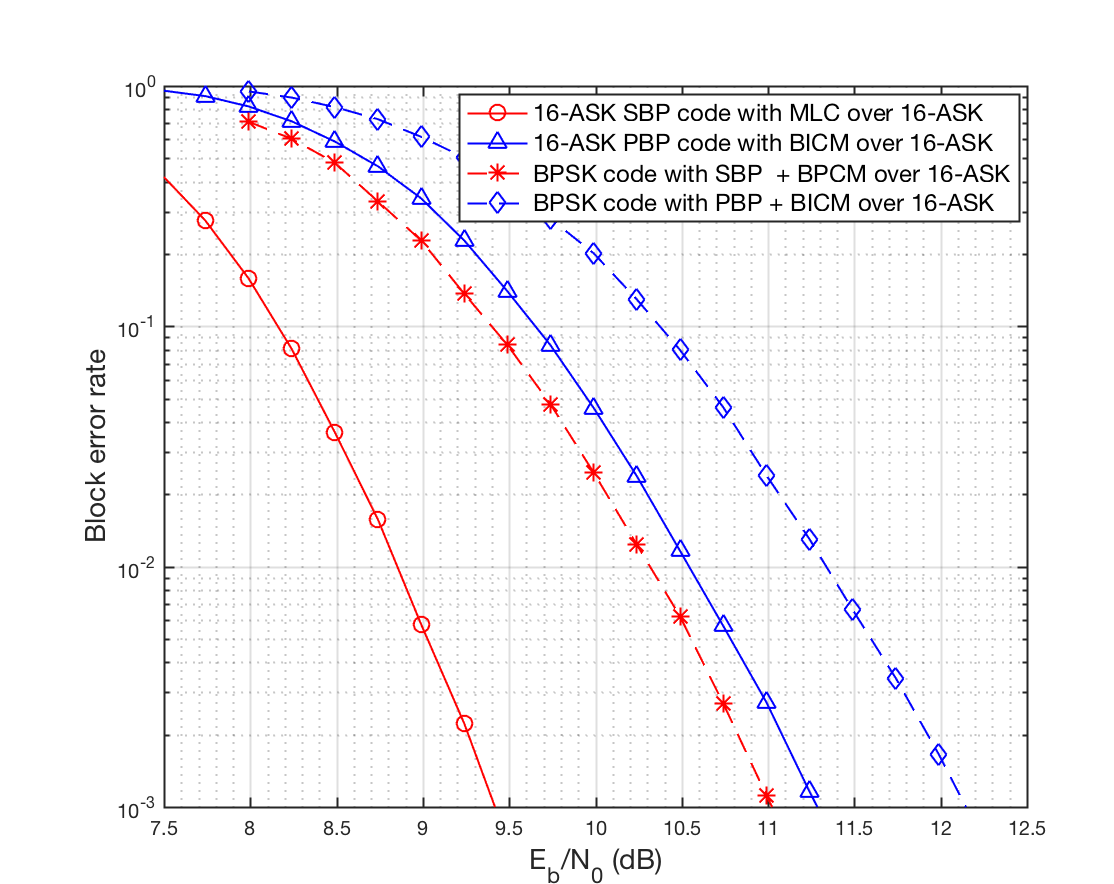}
\caption{16-ASK AWGN link sim; N=1024, rate=$\frac{1}{2}$.}\label{fig:16ask_bler_new}
\vspace*{0.1in}
\end{figure}

\begin{figure}[!ht]
\vspace*{-0.1in}
\includegraphics[width=\linewidth]{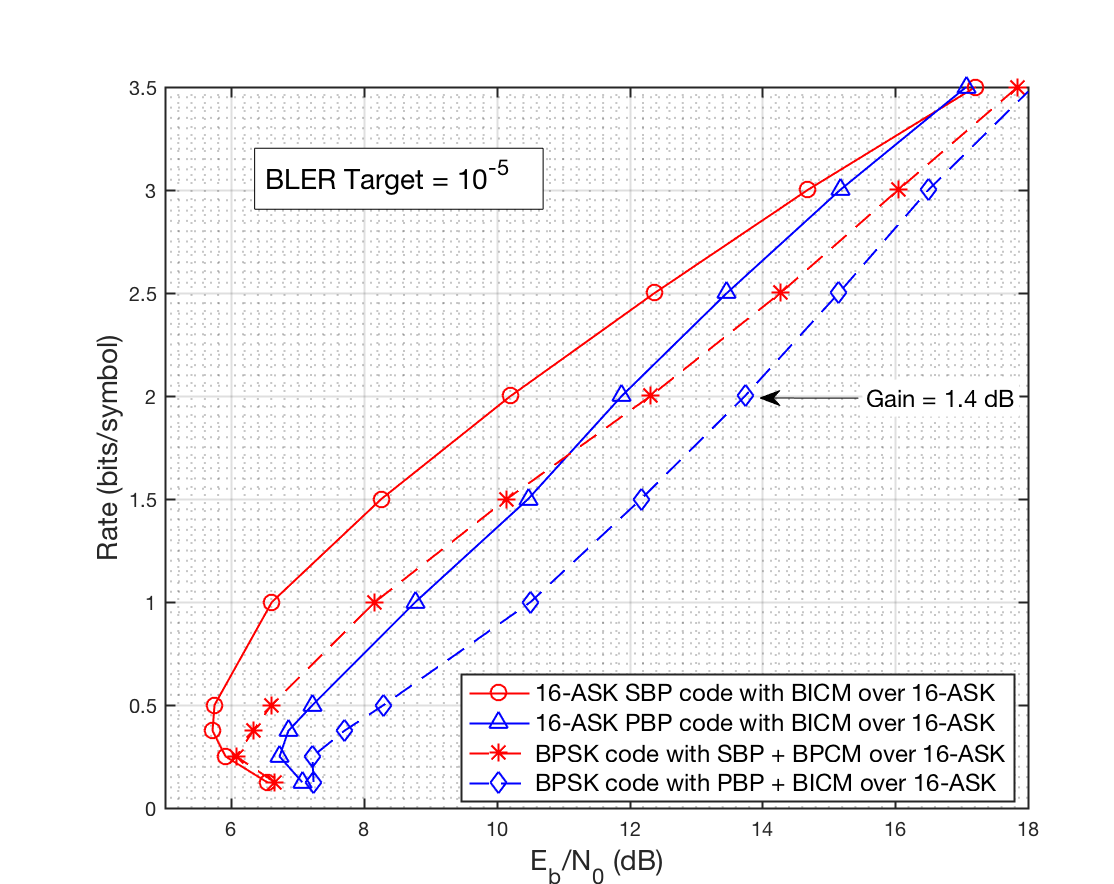}
\caption{16-ASK AWGN GA-DE; N = 1024.}\label{fig:16ask_de_new}
\end{figure}

The simulation setup is same as \hyperref[sec:4ask]{Section~\ref*{sec:4ask}}. The exact same BPSK codes are being used over 16-ASK. 
\hyperref[fig:16ask_bler_new]{Figure~\ref*{fig:16ask_bler_new}} shows AWGN link simulation results. \hyperref[fig:16ask_de_new]{Figure~\ref*{fig:16ask_de_new}} shows AWGN GA-DE results. 

The main results in the rate regime of interest are: 
\begin{itemize}
\setlength\itemsep{0.25em}
\item For BPSK codes, BPCM bests BICM by up to $1.4$ dB;
\item BPSK codes with BPCM is worse than modulation specific SBP code by up to $2$ dB;
\item BPSK codes with BPCM is worse than modulation specific PBP code by up to $1$ dB. 
\end{itemize}

\subsection{Design and results for 8-ASK}\label{sec:8ask}

For 8-ASK, number of bit levels is not a power of 2.  Prior work in \cite{Seidl} and \cite{HM} has suggested use of a special $3 \times 3$ polarization kernel either at every step or only the first step of polarization. In the context of modulation independent codes, changing the polarization kernel is not desirable. Here we suggest an alternate approach which is use number of subcodes to be 4 which is the closest power of 2 larger the number of bit levels. 

\begin{figure}[!ht]
\includegraphics[width=\linewidth]{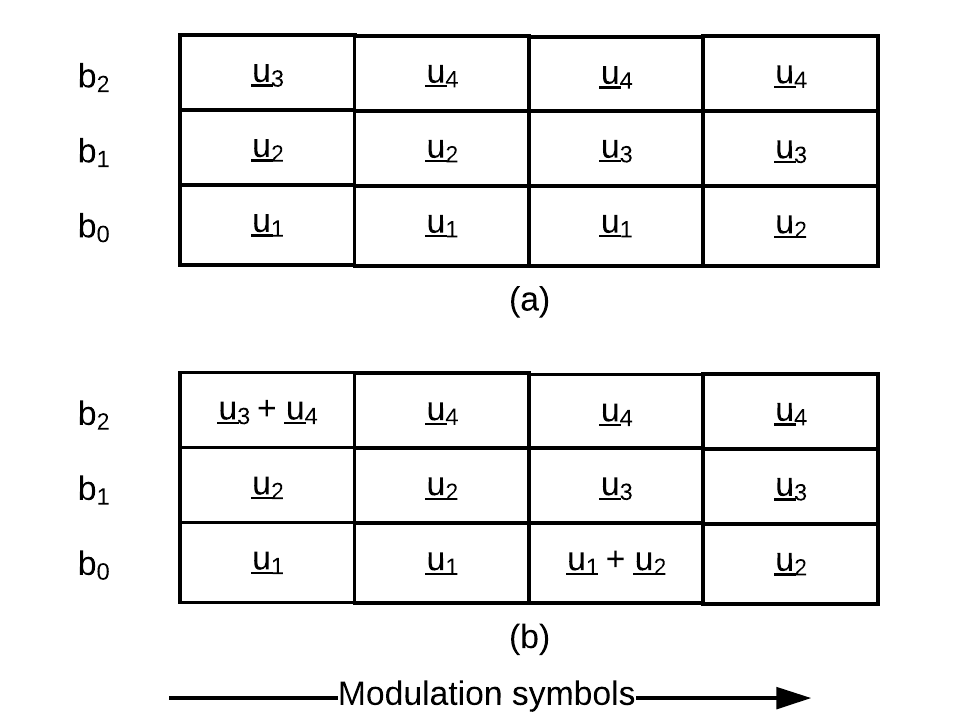}
\caption{8-ASK mapping}\label{fig:8ask_mapping}
\vspace*{0.05in}
\end{figure}

This is shown in \hyperref[fig:8ask_mapping]{Figure~\ref*{fig:8ask_mapping}} where $\underline{u_i}$'s show the 4 subcodes as described in \hyperref[sec:16ask]{Section~\ref*{sec:16ask}}. The \hyperref[fig:8ask_mapping]{Figure~\ref*{fig:8ask_mapping}}-(a) shows that the first subcode $\underline{u_1}$ mapped to the lowest bit level while the second subcode $\underline{u_2}$ is mapped partially to the lowest bit level and partially to the  middle bit level. The \hyperref[fig:8ask_mapping]{Figure~\ref*{fig:8ask_mapping}}-(b) shows a further optimization where selective polarization is applied to provide additional coding gain. The selective polarization:
\begin{itemize}
\setlength\itemsep{0.25em}
\item is applied within the same bit level;
\item is applied for code bits with the same index in the two subcodes, and is consistent with the polar transformation. 
\end{itemize}
Further when a subcode is mapped to two distinct bit levels, the code bits mapped to a given bit level are selected uniformly across the block length of the subcode. Finally, the way selective polarization is applied as shown in \hyperref[fig:8ask_mapping]{Figure~\ref*{fig:8ask_mapping}}-(b) allows for an MLC receiver across the 4 sub-codes of the polar code and the 3 bit-levels of 8-ASK. In addition to these change, $M_k = 8$ permutations used for \hyperref[alg:bpcm]{Algorithm~\ref*{alg:bpcm}}

\begin{figure}[!ht]
\vspace*{-0.1in}
\includegraphics[width=\linewidth]{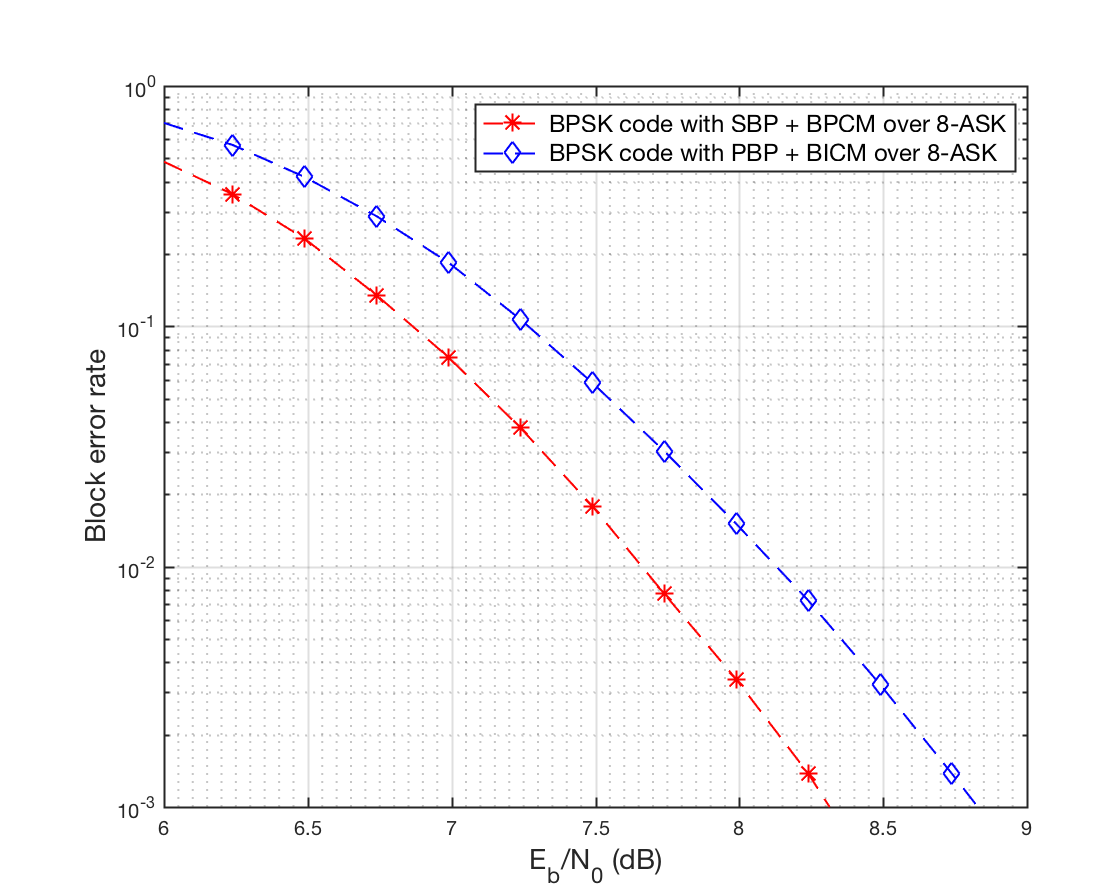}
\caption{8-ASK AWGN link sim; N=1024, rate=$\frac{1}{2}$.}\label{fig:8ask_bler_new}
\vspace*{0.05in}
\end{figure}

\begin{figure}[!ht]
\vspace*{-0.2in}
\includegraphics[width=\linewidth]{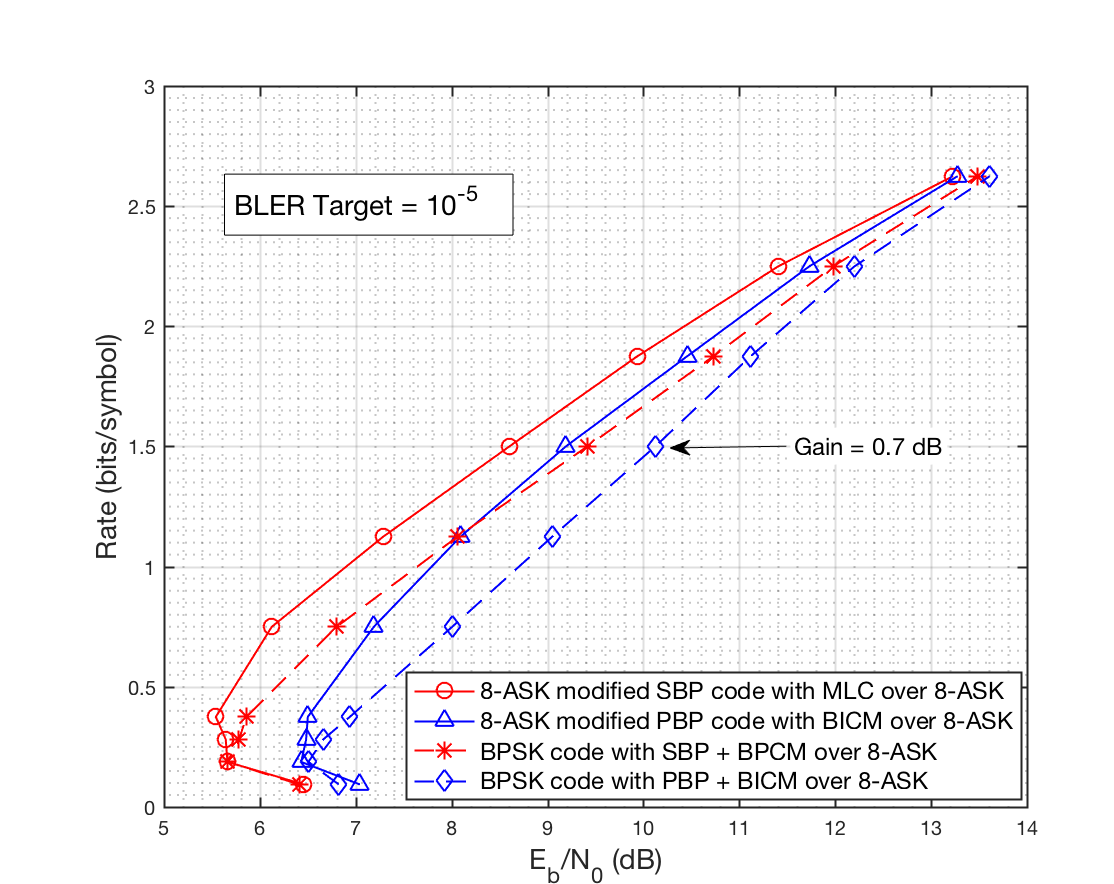}
\caption{8-ASK AWGN GA-DE; N = 1024.}\label{fig:8ask_de_new}
\vspace*{0.05in}
\end{figure}

\hyperref[fig:8ask_bler_new]{Figure~\ref*{fig:8ask_bler_new}} and \hyperref[fig:8ask_de_new]{Figure~\ref*{fig:8ask_de_new}} show the simulation results for the construction in \hyperref[fig:8ask_mapping]{Figure~\ref*{fig:8ask_mapping}}-(b). We note that the same construction of \hyperref[fig:8ask_mapping]{Figure~\ref*{fig:8ask_mapping}}-(b) is used for modulation specific SBP code construction for the results shown in \hyperref[fig:8ask_de_new]{Figure~\ref*{fig:8ask_de_new}}. 

The main results in the rate regime of interest are: 
\begin{itemize}
\setlength\itemsep{0.25em}
\item For BPSK codes, BPCM bests BICM by up to $0.7$ dB;
\item BPSK codes with BPCM is worse than modulation specific SBP code by up to $1$ dB;
\item BPSK codes with BPCM is worse than modulation specific PBP code by up to $0.2$ dB.
\end{itemize}

\section{Implementation aspects for BPCM}\label{sec:misc}

\subsection{Implementation complexity: MLC/BPCM vs BICM}

One of the benefits of BICM traditionally has been ability for the receiver to be able to compute bit LLR values in parallel for each bit level. This can translate to faster demodulator and decoder design. For Polar codes, however, the known decoding algorithms with good performance are successive cancellation decoders. For these decoders, the gains of parallel processing of bit LLRs is limited. Further in terms of number of computations, the MLC or BPCM approach would involve fewer number of computations compared to an accurate LLR computation for the BICM approach.

\subsection{HARQ with BPCM}

H-ARQ has been associated with the BICM approach traditionally due to the ease of storing LLRs per code bit. When using H-ARQ with an MLC or BPCM system, there would be two key challeges for the receiver:
\begin{itemize}
\item storing the received symbols (e.g. I/Q samples for wireless systems) rather than bit LLRs; 
\item more challenging soft combining since different modulations can be used across different transmissions.
\end{itemize}
These added complexities can be justified based on the performance gains observed for the MLC or BPCM approach.

\subsection{List decoding for BPCM}

In this paper, we have used the SCD algorithm. The BPCM/MLC architecture, however, can be used for list decoder \cite{Tal} as well. For BPCM/MLC architectures from the perspective of the list decoder is that the code should be viewed as one code rather than several shorter subcodes. For example, one CRC can be used across all the subcodes.

\begin{appendices}
\section{Interleaver design for BICM}\label{app:bicm}
\begin{figure}[!ht]
\vspace*{-0.15in}
\includegraphics[width=\linewidth]{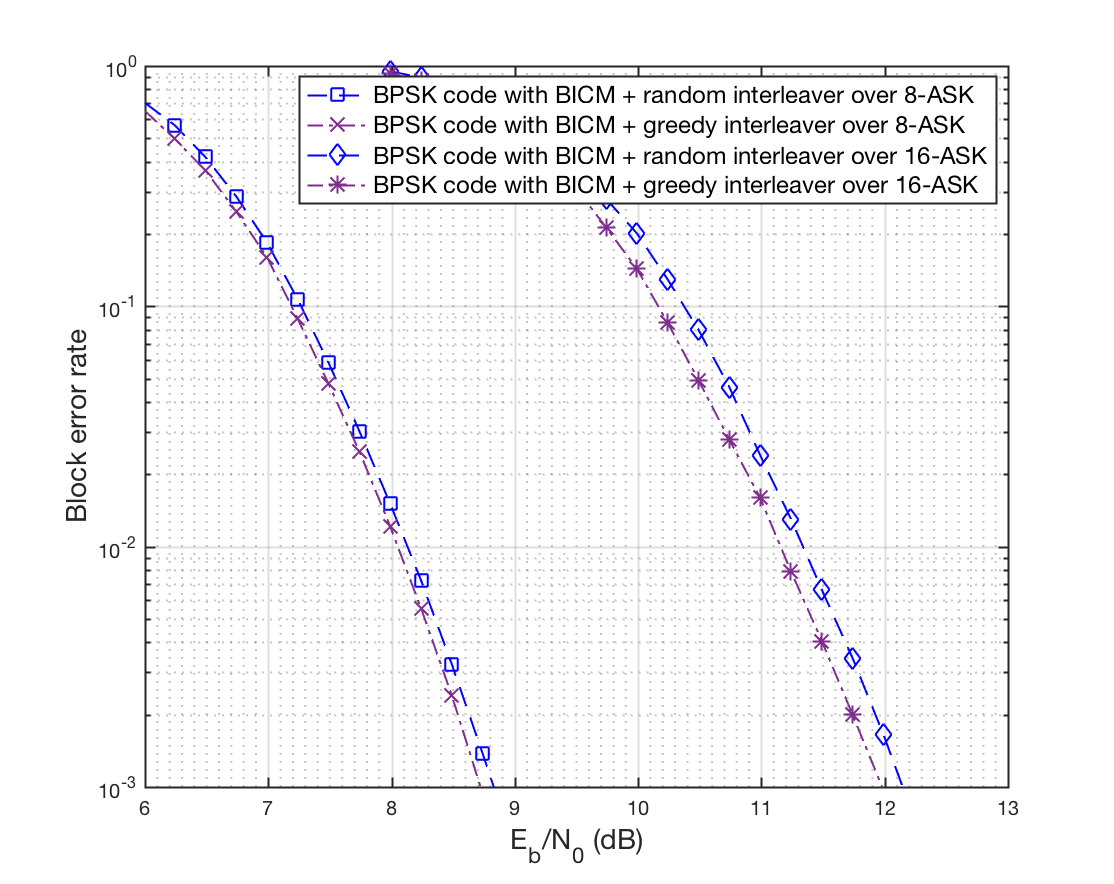}
\caption{8-ASK and 16-ASK AWGN link sim; N=1024, rate = $\frac{1}{2}$.}\label{fig:8_16_ask_bicm_bler}
\end{figure}

Performance of BICM can be improved by a well designed interleaver. See \cite{RY} and \cite{Alvarado} for interleaver design for Turbo and convolutional codes. Interleaver design for Polar codes is discussed in \cite{SLY} where the polar code is re-designed based on the interleaver. We assume that the polar code is not redesigned, and consider the approach of a greedy interleaver. In the greedy approach, we start with a random interleaver and swap two indices if swapping them would reduce the BLER estimate. This done over all $N*(N-1)$ choices of index pairs sequentially. The results shown in \hyperref[fig:8_16_ask_bicm_bler]{Figure~\ref*{fig:8_16_ask_bicm_bler}} show  a 0.1 dB gain for 8-ASK and a 0.2 dB gain for 16-ASK. A significant gain was not seen for 4-ASK.
\end{appendices}
\end{document}